\documentclass[10pt, a4paper, twocolumn]{article}
\usepackage{fixltx2e}
\usepackage{hyperref}

\usepackage[T1]{fontenc}
\usepackage[utf8]{inputenc}
\usepackage{authblk}
\usepackage{graphicx}
\usepackage{bm}
\usepackage[margin=0.5in]{geometry}
\usepackage[superscript]{cite}

\usepackage{titlesec}

\titleformat{\subsection}[runin]{}{}{}{}[]
\title{Communication and information processing in magnetic nanostructures with edge spin waves}

\author[1]{Antonio Lara}
\author[1]{Javier Robledo Moreno}
\author[2] {Konstantin Y. Guslienko}
\author[1]{Farkhad G. Aliev*}
\affil[1]{Dpto. Física Materia Condensada, C03, IFIMAC and INC, Universidad Autónoma de Madrid, 28049, Madrid, Spain}
\affil[2] {Dpto. Física de Materiales, Universidad del País Vasco, UPV/EHU, 20018 San Sebastián, Spain and
IKERBASQUE, the Basque Foundation for Science, 48013 Bilbao, Spain}

\date{\today}
\begin{document}
% Hint: \title{what ever}, \author{who care} and \date{when ever} could stand
% before or after the \begin{document} command
% BUT the \maketitle command MUST come AFTER the \begin{document} command!
%\maketitle

\twocolumn[
\begin{@twocolumnfalse}
\maketitle
*Correspondence and requests for materials should be addressed to F.G.A. (farkhad.aliev@uam.es)
\begin{abstract}

Low dissipation data processing with spins is one of the promising directions for future information and communication technologies. Despite a significant progress, the available magnonic devices are not broadband yet and have restricted capabilities to redirect spin waves. Here we propose a breakthrough approach to the spin wave manipulation in patterned magnetic nanostructures with unmatched characteristics, which exploits spin waves analogous to edge waves propagating along a water-wall boundary. Using theory, micromagnetic simulations and experiment we investigate spin waves propagating along the edges in magnetic structures, under an in-plane DC magnetic field inclined with respect to the edge. The proposed edge spin waves overcome important challenges faced by previous technologies such as the manipulation of the spin wave propagation direction, and they substantially improve the capability of transmitting information at frequencies exceeding 10 GHz. The concept of the edge spin waves allows to design broad range of logic devices such as splitters, interferometers, or edge spin wave transistors with unprecedented characteristics and potentially strong impact on information technologies.

\end{abstract}
\end{@twocolumnfalse}
]

\bigskip

Recently there has been an increasing interest in the use of spin waves for information transmission and processing \cite{corner1y5, corner2, corner3, corner4, corner4a,Lisenkov2016,Wang2017,diodex}. In contrast to electronics, the information encoded in the amplitude and phase of spin waves eliminates the contribution of charge dissipation. As a consequence, spin waves (SWs) propagate coherently through large distances even at room temperature. Contrary to the classic electromagnetic waves, a relatively small group velocity reduces the SWs wavelength down to the nanoscale. This opens new possibilities for a higher integration density in spin-wave logic devices through the effective manipulation of SWs propagating through periodic interferences or arrays of scatterers. The related research field of magnonics has been reviewed recently \cite{corner1y5, corner7, chumak, Hoffmann2015}. Gaining an effective control over SWs through their redirection \cite{corner8, corner9, corner10}, multiplexing \cite{corner11} or data processing \cite{corner4} is a very active research area nowadays. The global aim is the creation of SW based logic devices.\\
One of the most important features for future spin wave logics is the broadband capability of its building blocks. Also crucial is SW guidance, including the ability to split spin waves and/or redirect them at large effective angles. So far, the basic building blocks for SW transmission, redirection and control have consisted in straight or curved ferromagnetic strips \cite{corner8, corner10, corner11}. Such elements transmit information via SW modes which have a restricted frequency range (typically below 10 GHz) and are essentially suppressed after changing the direction of propagation up to 90 degrees or by increasing the drive frequency \cite{corner8, corner10, corner11}. This strongly limits the development of redirectional SW logic devices with broadband microwave capabilities.\\
One way to circumvent the problem is reprogrammable magnonic wave guides made out of dipole interacting elements with controlled switching of their local magnetization \cite{haldar}. A fundamentally different approach would be to create low dimensional SW logics with spin waves propagating along artificial or natural magnetic domain walls. Predicted by Winter \cite{Winter1961} SWs propagating along domain walls (DWs) have been identified in a number of recent experimental magnetization configurations in patterned films, involving the domain walls connecting magnetic vortices and edge magnetic charges \cite{stoll, perzlmaier, DMV, wagner}. The main drawback of the classic DW magnonics is its relatively low frequency of operation and certain difficulties to redirect SWs.\newline
$\quad$\newline
This work explores edge spin waves (E-SWs) propagating along edge domain walls (E-DWs) localized at magnetic nanostructure edges (triangular dot, see Fig. 1) and displaying a number of remarkable and unprecedented features. Besides the broadband character (with the capability to process information at frequencies above 10 GHz) E-SWs (i) can split easily comparing to the conventional spin waves, (ii) could be redirected at angles up to 120 degrees and (iii) show interference phenomena providing the possibility to create phase controlled nanoscale microwave emitters.\newline
$\quad$\newline
The proposed approach based on E-SWs could constitute a fruitful platform for the creation of low dissipation, ultrafast E-SW information processing devices such as spin wave diodes, multiplexors or phase detectors.  An effective incorporation of the spin waves in data processing devices requires their transmission and direction control using more complicated structures than just straight or curved micro waveguides - currently among the basic elements of magnonic waveguides \cite{corner3, corner4}.  The capability of changing the SW direction opens perspectives for the creation of novel logic elements based on the SW interference \cite{corner10}.

\section*{Results}
\subsection*{\textbf{Spin wave guidance along the edges.}}

\begin{figure}[h]
  \includegraphics[width=\linewidth]{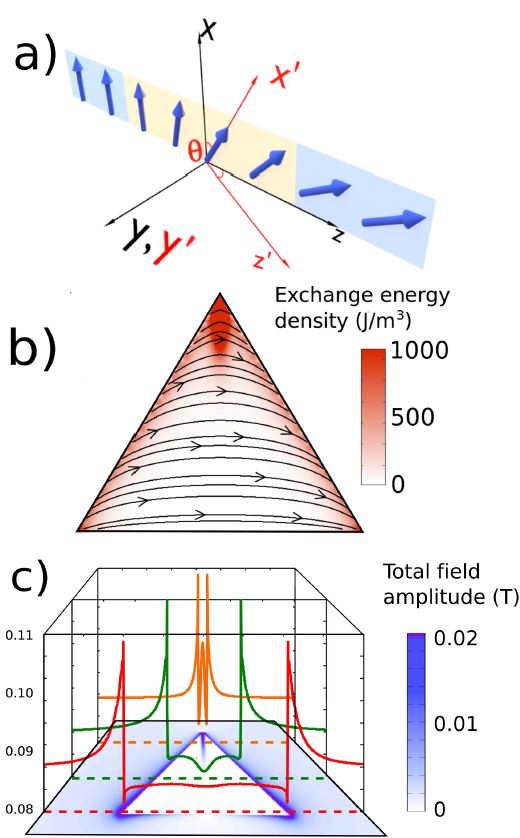}
  \caption{a) Right: N{\'e}el like domain wall. The yellow plane represents the domain wall, while the blue planes represent the domains separated by it. b) Exchange energy density profile (color scale) and magnetization distribution (streamlines and arrows) inside the dot. c) Total magnetic field inside permalloy triangular dot and cross sections at the places marked with dashed lines. In both cases the applied field $B=0.1$ T is parallel to the triangle base}
  \label{fig:corner_1}
\end{figure}

We consider patterned magnetic elements - equilateral triangular magnetic dots with the sides of 2000 nm and thickness 30 nm (Figure \ref{fig:corner_1}a,b). The static bias magnetic field $H_{DC}$ is directed along the triangle base. This geometry avoids butterfly-type E-DWs which appear when the field is applied perpendicularly to the base of triangle (so-called Y-state) and rectangles. Previous studies of the edge SWs carried out in rectangular and strip magnetic elements only used in-plane driving RF fields perpendicularly to the applied DC field \cite{jorzick, bls_center}. Therefore, increasing the DC field perpendicular to the edge affects E-DW related energy so that mainly inner strip SWs are excited above some critical field  \cite{Maranville2006,Demidov2008}.

In contrast to those studies, E-SWs in the triangular dots can be excited using a parallel pumping scheme (with homogeneous driving microwave magnetic field being parallel to the field $H_{DC}$) \cite{triangulos_jap} to avoid excitation of the quasi-uniform mode inside the device. Another important differentiating factor is that, as we shall see below, for triangular dots the edge modes could be excited locally, close to the dot edge, facilitating their propagation along the edges.

Indeed, as shown by the dashed lines (cross sections in Figure \ref{fig:corner_1} c), under a DC field parallel to one of their sides, the internal magnetic field distribution has minima near the dot edges which should provide localization of SW at the dot edges. The magnetization dynamics is described by the Landau–Lifshits–Gilbert (LLG) equation of motion governed by an effective magnetic field. SW guidance along E-DWs occurs because the effective field ($H_{eff}$) is formed by both the external and the internal magnetic fields, with the latter generated by the volume and surface magnetic charges.

So, increasing the properly aligned DC bias field under excitation localised near vertices just further confines the E-SWs towards the dot edges allowing a continuous field tuning of the E-SW frequency without breaking the SW topology. Our paper presents studies of two related cases: triangular dots in the so-called B state (the bias magnetic field is parallel to the triangle base) and the rectangular dots in the inclined external magnetic field.

\subsection*{\textbf{Analytical model for the edge guided spin waves.}}

Below we formulate an analytical model of the localized spin waves which are guided along an edge domain wall. The main idea is to consider the spin wave propagation in an inhomogeneous potential formed by the dot edge. We start by discussing the similarities and differences between the present model and Winter's model of SWs localized at the Bloch wall in an infinite ferromagnet \cite{Winter1961}.

Winter \cite{Winter1961} considered 180-degree Bloch DW connecting two semi-infinite oppositely magnetized ferromagnetic domains in an infinite ferromagnet with uniaxial magnetic anisotropy. Firstly, in our case instead of the uniaxial anisotropy constant ($K$) responsible for the domain formation in the Winter model, we have a dot shape anisotropy of magnetostatic origin.  The magnetization vector is parallel to the edge plane to minimize the side surface magnetic energy and then, it gradually rotates in the film plane to the direction parallel to the triangle base due to influence of the bias magnetic field applied parallel to the base. It has a maximum magnetization tilting angle of about 60 degrees due to the restricted geometry of the dot. In other words, the magnetization inhomogeneity near the dot lateral edges can be described as an incomplete Neel domain wall as sketched in Figure \ref{fig:corner_1}a). The magnetization rotates in the element plane to avoid face magnetic charges, see in Figure 1b. We consider that the dot edge plane (E-DW) plays a role similar to the DW plane in the Winter model and edge localized SWs propagating along the dot edge in the present case are similar to the Winter magnons propagating in the domain wall plane.

Secondly, in our case the E-DWs are naturally pinned by the element edge, providing (for soft magnetic permalloy patterned nanostructures) a dominant contribution to the additional magnetic anisotropy responsible for this pinning. On the contrary, due to the absence of such a natural edge pinning in the bulk ferromagnet Winter had to introduce an artificial pinning described by the additional magnetic anisotropy $K'$ \cite{Winter1961}. The stiffness of the Winter DW being proportional to the anisotropy constant $K'$ is assumed to be much smaller than the E-DW stiffness in triangular dot which is formed by the magnetostatic energy.

The magnetic energy of the dot can be described by the functional:

\begin{equation}
W[\bm{m}]=\int dVw(\bm{m})
\end{equation}

with the energy density:

\begin{equation}
w=A(\nabla \bm{m})^2-\frac{1}{2}M_s \bm{m}\cdot\bm{H}_m-M_s\bm{mH}
\end{equation}

where $A$ is the isotropic exchange stiffness constant, $\bm{m}=\bm{M}/M_s$ is the unit
magnetization vector, $M_s$ is the saturation magnetization, $\bm{H}_m$ is the magnetostatic field,
and $\bm{H}$ is external magnetic field (directed along the triangular base).

Using Eq. (2) we get the dispersion relation for the edge localized spin waves in triangular
dots as the following (see Methods):

\begin{equation}
\omega^2(\kappa)=\omega_M^2[1+l_e^2 \kappa^2 +(1-\nu^2)h][l_e^2 \kappa^2+(1-\nu^2)h] ,
\end{equation}

where $\kappa$ is the spin wave vector along the dot edge, $l_e=\sqrt{A/2\pi M_s^2}$ is the exchange length, $h=H/4\pi M_s$ is the reduced magnetic field
parallel to the triangle base, and $\nu=\xi l_e/\sqrt{h}$ is a parameter of
the model. The parameter $\nu$ is the ratio of the spatial decay of the dynamic and static
magnetizations increasing the distance from the dot edge. The calculation of the value
of $\nu$ is beyond the present approximate model. We can state only that $\nu<1$ to secure
a frequency increase with $h$ increasing.
The domain wall considered above is stable in the bias field $h>h_c$ , where the
critical field $h_c$ corresponds to transition from the dot vortex state to B-state increasing
the field $h$ directed along the triangular dot base. The frequency gap $\omega(\kappa=0)$ is due to
a finite value of the bias magnetic field $h$ forming the Neel wall.
We note that all static and dynamic magnetostatic energies (except the energy of
the face magnetic charges) were neglected deriving Eq. (3). Therefore, it is valid within
the limit of thin dots, when the dot aspect ratio thickness/(in-plane size), $t/a<<1$.

\subsection*{\textbf{Local excitation and control of edge spin waves.}}

In the following we describe results on numerical experiments which support the possibility to excite, redirect and operate E-SWs and to fabricate E-SW magnonics devices. The presence of both surface and point (vertex) magnetic charges allows effective channeling,  re-routing and external control of an unprecedented broadband frequency E-SWs.

\begin{figure}[h]
  \includegraphics[width=\linewidth]{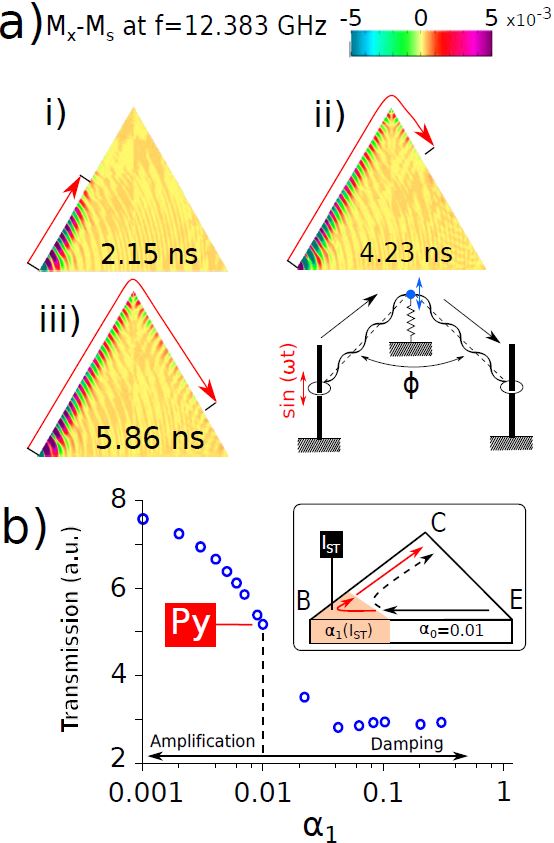}
  \caption{a) Snapshots of normalized x component of the dynamical magnetization in a 2 $\mu$m long, 30 nm thick triangular dot at different moments after a continuous excitation of 12.383 GHz is applied to the bottom left corner. The right bottom figure sketches a mechanical analogy for the excitation and propagation of edge spin waves. b) Simulated transmission of E-SWs from the bottom left  corner (E) to the right corner (C) vs. local damping at the top corner (B). Spin current $I_{ST}$ should provide local control over the damping $\alpha_1(I_{ST})$.}
  \label{fig:bend_corner}
\end{figure}

Based on this fact, one can think of using the edges as transmitters of spin waves, generated at the corners of the triangles. To test this idea we have performed simulations in which a DC film saturates the sample to achieve the B magnetization state (Figure \ref{fig:corner_1}b) and a local AC magnetic field applied only at the base linked vertex of the triangle. The latter generates spin waves that propagate with a considerable amplitude along the dot edges, exceeding the rest of the sample (Figure \ref{fig:bend_corner}a) ). The top corner behaves as a second emitter when the E-SWs reach it, and the waves start propagating along the second ``channel'', along the other lateral edge of the dot. The net effect is that waves are transmitted through the corner without reflections and therefore change their direction without any significant loss of intensity. Of course, as they advance through the ferromagnetic material they are damped, and their intensity decays with distance, but the presence of the corner does not yield any additional loss. With this method, E-SW emitted at one corner (for example, at bottom left corner) can reach the opposite corner (bottom right corner), not in a straight line, but through a longer path (bottom left - top - bottom right corner). Then, E-SWs prove that they can be used to overcome a deflection angle of (at least $120^{\circ}$). Our simulations show that the present strategy works also for isosceles triangles.

The right bottom part in Figure \ref{fig:bend_corner}a presents a mechanical model explaining the redirection mechanism behind the E-SWs. Two strings are connected at point through so-called Robin boundary conditions (i.e. mass / massless point fixed by spring) represent E-DWs inter-connected at the upper triangle vertex by the magnetic charge  (Fig. \ref{fig:corner_1}c). String waves (E-SWs) could be in principle redirected within a large range of angles $ (0<\Phi < \pi ) $ if two strings are kept tense  (i.e. E-DWs are present) and are interconnected through Robin conditions  (vertex magnetic charge).

We stress that not all E-SW frequencies are equally valid for being transmitted between the vertices. The spin waves eigenfrequencies are determined by the applied field, which in turn controls the width of the channels. The physical size of the dots is not a factor that affects these eigenfrequencies, as long as we are comparing regions without bends, corners, etc. In the straight sections of the edge, as we shall see further below, only the applied magnetic field affects the wave frequency.

The set consists of around 10 eigenfrequencies that are typically available in the triangular device (Figure \ref{inclined}).  Edge SWs with the lowest frequencies ``see'' a very wide ``exchange channel'', while waves with the largest frequencies ``see'' how it narrows until, eventually, the ``channel'' is too narrow for the waves to propagate through. We have found that waves with intermediate frequencies show an optimum relation between their wavefront width (too wide wavefronts do not result in waves propagating further than the corner, due to interferences between the incident and deflected waves) and their amplitude to be used in devices.
The simulations show that E-SWs are robust to variations of the triangle dimensions between 1 and 2 $\mu$m, as well as to the triangle aspect ratio.

The possibility of E-SWs that propagate through the dot corners opens perspectives for creation of a new class of reliable and relatively simple information processing devices. To start with, one can imagine a number of ways to interrupt the E-SW propagation leading to E-SW diode. This could be done, for example by limited rotation of DC magnetic field.  Figure \ref{fig:inclined} demonstrates how the exchange energy distribution varies with angle of the DC field. Strong sensitivity of the E-SW spectra to the DC field angle allows the interruption of the E-SW propagation through small changes in the DC field angle, as well as to make the spectrum of SWs change.

Another (easier to realize on the small scales) strategy of blocking E-SW propagation from E-SW emitter ($E$) to collector ($C$) could be through injection of spin polarized current to modify locally the damping close to an upper vertex with strongest magnetic charge (base $B$, see sketch in inset to Figure \ref{fig:bend_corner}). High damping should result in an effective blockade of the waves before they turn the upper corner $B$, therefore never reaching the right bottom corner $C$. Inset to Figure \ref{fig:bend_corner} shows the simulated transmitted E-SW power when emitted from the bottom left corner to the bottom right corner, passing through the upper one, and with a variable damping $\alpha_1$ close to $C$. While for the locally suppressed damping the transmitted power is substantially increased for $\alpha_1<\alpha_0$ ($\alpha_0$ is damping of Permalloy). In the opposite limit with $\alpha_0\simeq1$ one observes a plateau with even some small increase of the transmitted power caused by bouncing back of the E-SWs without entering to the high damping region (Figure \ref{fig:bend_corner})b).

\subsection*{\textbf{Edge spin waves interferometer and splitter.}}

\begin{figure}
  \includegraphics[width=0.9\linewidth]{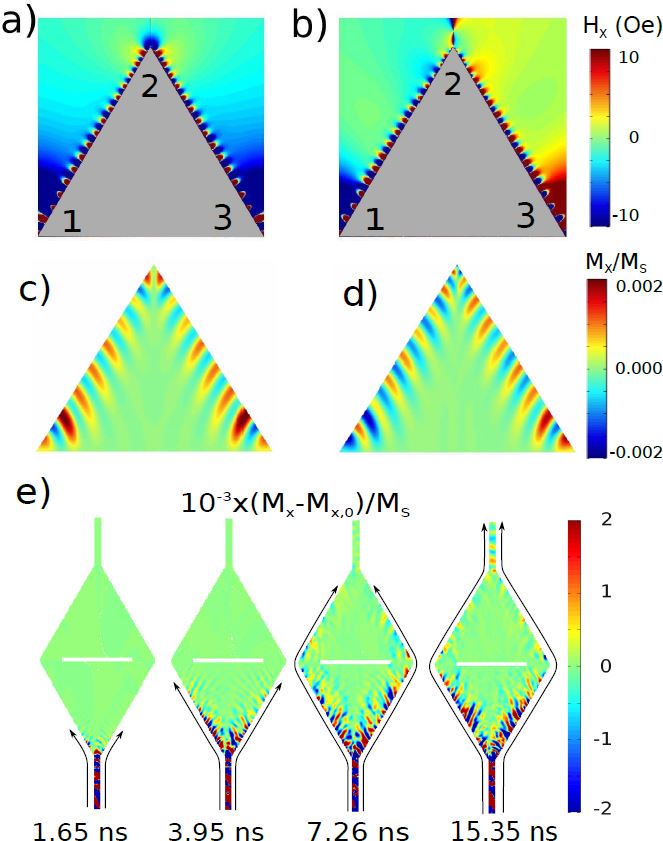}
  \centering
  \caption{Distribution of the magnetic field ($x$ component) outside of the triangular dot when applying a sinusoidal excitation in both bottom dot corners. In a), both corners are excited in phase. In b), the corners are excited with a 180º phase difference. c) and d) show the same as a) and b), but in the $x$ component of magnetization inside the dot. The excitation frequency is f=11.259 GHz. e) shows simulations of a spin wave splitter. The edge character of SWs allows to overcome obstacles for some frequencies, such as the slit in the middle. The excitation frequency is 12.73 GHz.}
  \label{fig:interference}
\end{figure}

The design of the E-SW interference device based on a single triangular dot is straightforward. For a constructive interference, two in-phase signals can be sent from the two bottom dot corners producing waves that propagate through their respective ``channels'', to end up interfering at the upper corner (Figure \ref{fig:interference}). They could be detected, for example, through  measurement of the local stray magnetic field near the upper corner. A minimum stray field is observed in the proximity of the upper corner in the conditions for destructive interference.
A precise control over the relative phase and amplitude of both emitters opens a wide variety of possibilities in terms of interfering waves.

The nanometer sized stray field tuned by E-SW interference in vertex of a single triangular dot could be used to create near field microwaves (NFM). For example, when in-phase E-SWs are excited in the dot corners (1,3), constructive interference results in the enlarging of the total oscillating local field near point (2) in the an about 10 nm proximity to the vertex. On the other hand, with the out-of-phase interference, one could create extremely localized oscillating magnetic fields in a few nm around corner (2) (Figs. \ref{fig:interference}a,b). Having gained control over NFM emission in the triangular dot device, one could further explore the NFM capabilities, for example, for effective writing bits with triangles as write heads in the hard disk drives, in substitution of the currently implemented inductive coils.

Since standard SW technology is based on the propagation and detection of SWs through the ferromagnetic strips, we can also consider a E-SW splitter and interferometer based on a strip with two connected by their bases triangles (Figure \ref{fig:interference}e ). The idea is to use a triangle-like structure in the B state to split effectively spin waves directed from a ferromagnetic strip. Figure \ref{fig:interference}e demonstrates by simulations the effective splitting of spin waves propagating along the strip onto two E-SWs at the intersection between strip and the triangular shape. The proposed E-SW splitter, analogously to the standard optic splitters, allows the realization of an effective interference through backwards transition in the point where opposite strip line situates (see Figure \ref{fig:interference}e). We have introduced a slit between two triangles to show that only E-SWs contribute to the backwards interference process.

\subsection*{\textbf{Generalization of edge spin waves to magnetic strip.}}

Next we demonstrate that edge spin waves are not specific to magnetic triangular dots only, but could be excited along edges of magnetic rectangular elements (strips) too when a quasi homogeneous DW is created along the device edge. In order to excite spin waves along the edges of a magnetic strip, contrary to the usual practice, the applied magnetic field should not be perpendicular to the strip direction, but tilted at an angle in order to eliminate the possibility of the ``butterfly type'' interruption in the centre of the exchange energy channels. For an arbitrary geometry in magnetic nanostructures (as long as they are not too thin, which would force all the magnetization to lay parallel to the edges), an external field which is inclined with respect to the edges results in uninterrupted exchange channels that can be used for propagating spin waves.

\begin{figure}
  \includegraphics[width=0.9\linewidth]{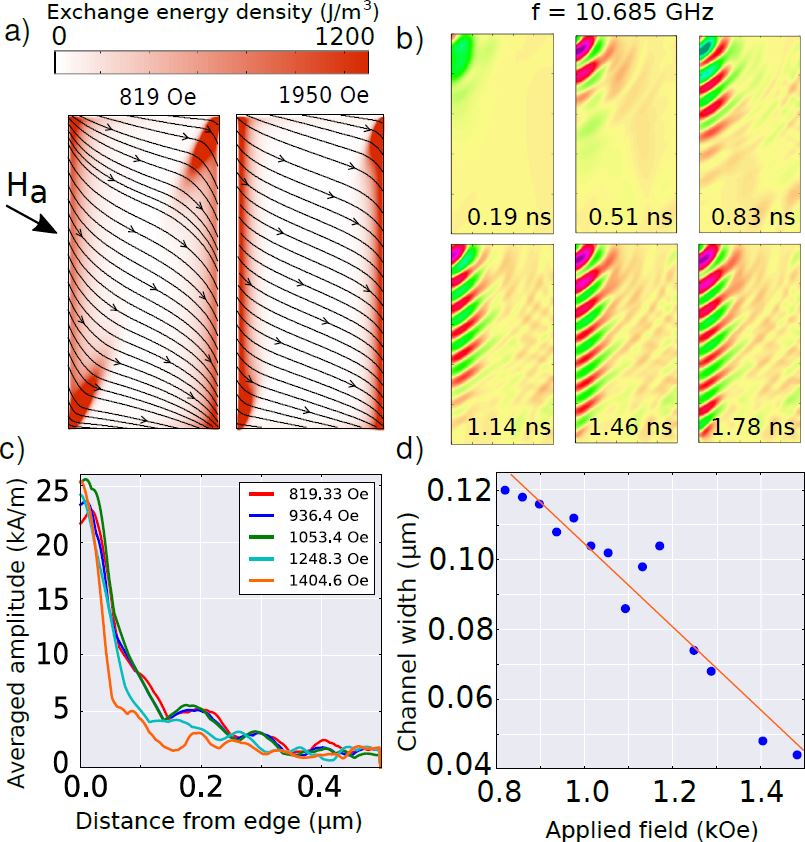}
  \centering
  \caption{Edge spin waves in rectangular elements: a) the exchange energy density; b) snapshots of the edge spin waves excited at the top left corner at an applied field 819 Oe, at 20$^\circ$ from the horizontal; c) amplitude of the edge spin waves (averaged over the vertical side of the rectangle) vs. distance from the edge; d) width at half height of the exchange channels vs. magnitude of the applied field at 20$^\circ$ from the horizontal.}
  \label{fig:rectangule_summary}
\end{figure}

Figure \ref{fig:rectangule_summary} summarizes our results on the generation of E-SWs in a 30 nm thick Permalloy strip with lateral dimensions of 1  by 0.5 $\mu$m. Our simulations show that for the chosen sample geometry the most homogeneous E-DW appears when an in plane magnetic field is inclined to about 20$^\circ$ with respect to the vertical to the long side of the strip. We also demonstrate effective propagation of broadband E-SWs in rectangular element  (Fig.\ref{fig:rectangule_summary}b). Finally, Figures \ref{fig:rectangule_summary}c,d show how the E-DW channel width decreases with the magnitude of inclined DC magnetic field.

\subsection*{\textbf{Comparison with between theory, simulations and experiment.}}

In order to excite mainly E-SWs in magnetic substructures one has to create as homogeneous as possible E-DW and predominantly excite the spin waves near DW extremes. We approach these conditions experimentally in the array of magnetostatically decoupled triangular dots and investigate the sample microwave response with microwave field parallel to the applied DC field. In that configuration we both create edge domain wall to channel Winter magnons (Figure \ref{fig:corner_1}) and excite them mainly locally close to vertices.

\begin{figure}[h]
\centering
\includegraphics[width=\linewidth]{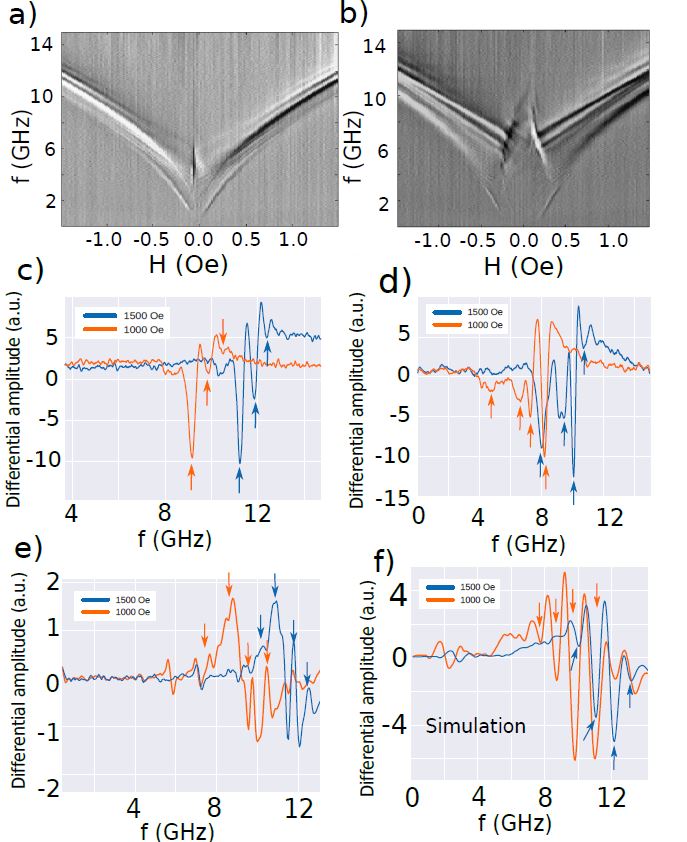}
\caption{Experimentally measured microwave permeability as a function of driving frequency and in-plane DC magnetic field for 2 $\mu$m triangular permalloy dots with thickness of a) 30 nm and b) 100 nm. c) and d) show two cross sections of the spectra a) and b) respectively. e) and f) show the simulated excitation spectra of a 2 $\mu m$ triangular dot (30 nm and 100 nm thick, respectively) for two different applied fields. Arrows indicate some of the strongest eigenmodes.}
\label{fig:meas_modes}
\end{figure}

Our results are shown in Figure \ref{fig:meas_modes}, where we compare the measured spectra of an array of 30 nm thick dots (part a) with an array of 100 nm thick dots (part b). All dots are triangular, with a lateral size of 2 $\mu$m. The interesting feature of these results is that it is clearly visible how the highest frequency branches are stronger at high fields, and almost invisible at low fields, where the lowest frequency modes are stronger. This confirms the relationship of the width of the exchange channels and E-SW bandwidth with the applied field as observed in simulations and experiment (Fig. \ref{fig:meas_modes}). Higher frequency waves can only propagate through narrow channels, achieved at high fields. Likewise, the intensity of the low frequency modes become weaker at high fields. Another interesting feature (which will be discussed in details elsewhere) is presence of the modes typical for the magnetic vortex state observed at low fields only. As expected, the vortex state exists in the broader field range for 100 nm thick dots in comparison with 30 nm thick dots.

\begin{figure}[h]
\centering
\includegraphics[width=\linewidth]{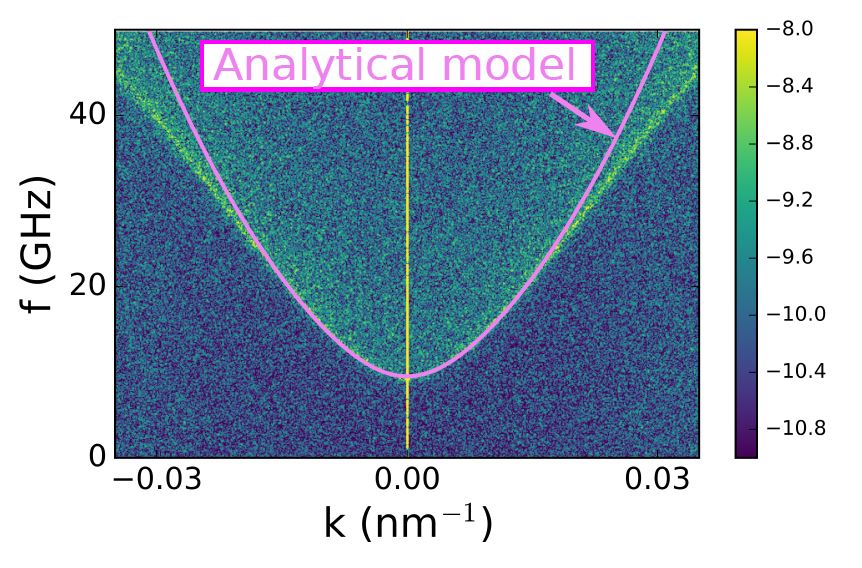}
\caption{Simulations (color scale of the 2D FFT amplitude, a.u.) and analytic calculation (line) of the dispersion relation of spin waves propagating in a semi-infinite plane with a 60$^\circ$ Neel domain wall localized at the boundary. The bias magnetic field is $H$=1000 Oe.}
\label{fig:dispersion_theory_simulations}
\end{figure}

In order to corroborate the main analytical result (equation 3) numerically, we have carried out simulations of the dispersion relation for the E-SWs modes excited by uniform microwave field in an infinitely long strip of 1 nm thickness, (with periodic boundary conditions in the $x$ direction, and not in $y$) subjected to an inclined in-plane DC field of magnitude 1000 Oe directed at 60$^\circ$ with respect to the sample face. This has been done because the analytic model considers E-SWs propagating along a semi-infinite plane, within a domain wall without surface magnetic charges. Thicker samples provide a more complicated dispersion curve, with several branches, not accounted by the analytic model, which is valid for very thin samples.
Figure \ref{fig:dispersion_theory_simulations} compares the E-SW frequencies as a function of the in-plane wave vector obtained numerically (colored 3D plot) and analytically by using Eq. (3) (line) with only one free parameter of the model tuned (we used $\nu =0.2$, the frequency is weakly dependent on $\nu $  up to 0.4).
Taking into consideration the above discussed applicability limits of the theory, we find the agreement to be satisfactory.

\subsection*{\textbf{Factors to be considered in real device fabrication.}}

The following factors should be considered during design of E-SW based real devices. Our simulations do not take into account possible influence of the edge related structural disorder.  The influence of the structural disorder could be two-fold. First, it could change locally the magnetic damping. The deviation from perfect symmetry of the upper vertex might be important to consider for optimized E-SW transmission between the triangular dot sides. While the above factors would be less important for few micron sized triangles, they should be more carefully considered in nanoscale triangles. One way to circumvent the edge related disorder would be by using self-assembled magnetic triangular dots, such as triangular (111)-oriented LSMO nano-islands \cite{Zabaleta2014}.

Further studies should also verify possible influence of coupling to the substrate on E-SWs. One of the routes to improve applicability and functioning of the E-SW based communication devices is fabrication them using YIG films instead of metallic Permalloy, taking advantage of dramatic reduction of the magnetization damping in these insulating ferromagnets. As to the need in application of the external magnetic field, its role in creation of the B-state could be played by exchange bias field with an underlying antiferromagnet. Figure \ref{fig:antiferro} shows simulations of the hysteresis loops of a triangular dot placed over a layer of fixed spins, which represent an antiferromagnet. Different exchange couplings between the two layers shift the hysteresis cycle to higher or lower fields, so that the B or Y dot magnetization state could be achieved at zero external field using this method.

\section*{Conclusions and perspectives}

In \textit{Conclusion}, the proposed here edge spin waves could potentially become a unique tool for low dissipation and broadband transmission and processing of information. Their unmatched features allow for effective control and redirection of the spin waves through topologically protected well defined and narrow paths. This concept can be used to design different spin wave logic and sensing elements, such as diodes, multiplexors, read/write heads etc. Also, modulation of the signals could be investigated in more detail in order to produce more complex, non linear devices, logic gates, mixers, etc. Engineering of the shapes of the nanostructures would allow to use edge spin waves to suit different needs, for specific systems of different sizes, configurations, etc. For larger systems, where the waves could be damped out before they reach their final destination, it is important to find materials with low magnetic damping. At present, YIG seems to be the ideal candidate, regardless some difficulties in its growth and lithography with small size devices, as compared to the metallic ferromagnets such as permalloy. Finally, we believe that some of the proposed strategies to develop topologically protected E-SW based communication and signal processing devices could also be implemented in other nanoscale structures. For example, spin-wave edge excitations have been predicted recently in the arrays of dipolarly interacting magnetic nanodots  \cite{Lisenkov2016} and ferromagnetic spins on a honeycomb lattice with perpendicular anisotropy \cite{Wang2017}.

\section*{Methods}
\footnotesize{
\subsection*{\textbf{Experimental set-up.}}

The high-frequency measurements were taken using a network analyzer (VNA), feeding a variable frequency microwave signal to a coplanar waveguide, onto which the sample is placed. The waveguide generates a variable frequency magnetic field which drives the sample. When a resonance frequency is excited, the sample absorbs more energy and we see a decrease in the VNA transmitted signal from the emission port to the detection port. An electromagnet provides the necessary external magnetic field to bias the sample. A computer automatizes the process of fixing a static magnetic field, and sweeping the high frequency, and finally save the spectrum before going to the next static field. For further detail refer to \cite{DMV}.

\subsection*{\textbf{Micromagnetic simulations.}}

The micromagnetic simulations were carried out using the OOMMF code. The typical magnetic parameters for Permalloy ($Ni_{80}Fe_{20}$ alloy) were assumed: saturation magnetization $4\pi M_S=10.43* 10^3 $A/m, damping constant $\alpha =0.01$, exchange stiffness $A=1.4*10^{-11}$ J/m. To find the eigenfrequencies of a magnetic structure, a $10^{-12}$ s gaussian in-plane magnetic field pulse was applied uniformly. After a 10 ns relaxation time, with the Fourier Transform of $M_x(t)$ magnetization component the spin wave eigenmode spectrum is obtained. Knowing the eigenfrequencies, which appear as peaks in the absorption spectrum, a local excitation at a single eigenfrequency can be applied to observe the response of the whole structure (i.e., to observe the propagation of spin waves away from the source), for as long as it may be necessary for the spin waves to reach the target area. To obtain the dispersion relation in figure \ref{fig:dispersion_theory_simulations}, an almost semi-infinite plane (the rectangular sample with periodic boundary conditions in one direction, and a large enough size in the other) is excited with a short field pulse, all the time under the influence of an inclined DC field at 60$^\circ$. The $z$ projection of the magnetization in the simulation cells parallel to the boundary  of the sample, and near it (to make sure we are considering waves in the edge domain wall) is saved for different times. With this space vs. time 2D array of data, a Fourier Transform in 2D provides the corresponding wavevector - frequency array, whose amplitude represents the dispersion relation of the waves along the selected path.

\subsection*{\textbf{Analytical calculations.}}

To conduct analytic calculations we introduced the coordinate systems, $xyz$ and $x'yz'$ for description of the magnetization and SW wave vector components. The axis $z$ is in the dot plane and perpendicular to the dot edge plane $x0y$, the $x$ axis is along the dot edge in the dot plane, and $y$ axis is perpendicular to the dot plane (see Fig. 1b). The angle $\theta$ of the dot in-plane static magnetization \textbf{$M_0$} changes from zero to 60 degrees. It is convenient to describe magnetization dynamics in the coordinate system related to the \textbf{$M_0$} direction denoted as $x'$ axis. In this coordinate system the dynamic magnetization components are $m_{z'}$ and $m_y$.

The static magnetization \textbf{$M_0$} is parallel to the dot edge (the plane $z=0$), $\theta(0)=0$,
due to the surface magnetostatic energy. Then, the angle $\theta(z)$ increases with the coordinate $z$
increasing due to finite bias magnetic field applied parallel to the triangular dot base (a) and forms a N{\'e}el like domain wall (Fig. 2). We account that the domain wall localization area is much smaller than the triangle side and consider the domain wall texture far from the triangle corners. We also assume that the triangle dot thickness $t$ is much smaller than the equilateral triangle size $a$. This assumption allows us to neglect all magnetostatic energies except the energy related to the face surface magnetic charges. It is convenient to introduce the angle $\phi(z)=\pi /3-\theta(z)$ . The static Lagrange-Euler equation for our geometry defining an equilibrium dependence of $\phi(z)$ is:

\begin{equation}
l_e^2\phi_{zz}=h\sin \phi .
\end{equation}

Eq. (3) admits the solution $\tan (\phi(z)/4)=\tan(\phi(0)/4)e^{-\sqrt{h}z/l_e}$, where
$\phi(0)=\pi/3$ . It describes a 60-degree Neel domain wall, where the magnetic field $h$ plays a role of
the uniaxial magnetic anisotropy responsible for the wall formation.

To consider the magnetization dynamics on the background of the 60-degree Neel domain wall introduced above we use the Landau-Lifshits equation of magnetization motion:

\begin{equation}
\frac{\partial \bm{M}}{\partial t}=-\gamma\bm{M}\times\bm{H}_{eff}, \quad  \bm{H}_{eff}=-\frac{\partial w}{\partial \bm{M}}
\end{equation}

It is convenient to use spherical angles of the magnetization vector $(\Theta, \Phi)$ in the
$xyz$ coordinate system and write the decomposition $\Theta=\Theta_0+\vartheta$, $\Phi=\Phi_0 + \psi$, where the spherical angles $(\Theta_0=\pi/2-\theta, \Phi_0=0)$ describe the static domain wall profile and the angles  $\vartheta, \psi$ describe small deviations from the equilibrium magnetization direction. Then, we search for the
solution in the form of plane spin waves propagating along the dot edge, $xOy$ plane, $\vartheta(x,z,t)=\vartheta(z,\omega)e^{i(\kappa x - \omega t)}$, $\mu(x,z,t)=\mu(z, \omega)e^{i(\kappa x - \omega t)}$, where $\mu=-\psi \sin(\Theta_0) $.
Any dependence of the static and dynamic magnetizations on the dot thickness
coordinate $y$ was neglected assuming thin enough dot. The system of equations of
motion for the spin wave profiles $\vartheta(z,\omega)$, $\mu(z, \omega)$ is the following:

\begin{equation}
-i\frac{\omega}{\omega_M}\vartheta=-\mu''+[1+l_e^2\kappa^2+h\cos \phi(z)]\mu
\end{equation}

\begin{equation}
i\frac{\omega}{\omega_M}\mu=-\vartheta''+[l_e^2\kappa^2+h\cos \phi(z)]\vartheta
\end{equation}

where $\omega_M = \gamma 4 \pi M_s$, $\phi(z)$ is the domain wall profile, and the spatial derivatives are taken with respect to the dimensionless coordinate $z/l_e$.

The function $\cos \phi(z)$ is a smooth increasing function of $z$ , varying from $1/2$ to $1$
increasing $z/l_e$ from 0 to infinity. It describes a potential well for the spin waves
traveling along the triangular dot edge. The well exists solely due to finite value of the bias magnetic field $h$. The bottom of the well is located at the dot edge $z=0$. Increase
of the value of the bias field $h$ results in decrease of the localization area of the spin waves near the
edge $z=0$. To find an approximate solution of the system of equations (5) we substitute the variable
term $h \cos \phi(z)$ to its spatial average over the distance $l>>l_e$, $h \cos \phi(z) \rightarrow \wedge{h}\approx h-O(l_e/l)$. Then, we search for the solution of Eq. (5) in the form of the edge localized spin waves $\vartheta(z,\omega)=\vartheta(\omega)e^{-\xi z}$, $\mu(z,\omega)=\mu(\omega)e^{-\xi z}$ and get a homogeneous system of equations for the spin wave amplitudes $\vartheta(\omega)$, $\mu(\omega)$. Calculating the determinant of this system and equaling it to zero yields the dispersion relation for the spin waves displayed in the main text.
}

%\bibliography{biblio}{}
%\bibliographystyle{IEEEtran}

\section*{Acknowledgements}

Authors acknowledge Jose Luis Prieto for samples fabrication. The work in Madrid has been supported by SVORTEX through CCC-UAM, by Spanish MINECO (MAT2015-66000-P) and the Comunidad de Madrid through NANOFRONTMAG-CM (S2013/MIT-2850) grants.

\section*{Author contributions}

F.G.A. supervised the project.  A.L. performed numerical simulations and experimental measurements with help from J.R.M.  The theoretical model has been developed by K.G.. The paper was written by F.G.A., A.L. and K.G. All authors commented on the paper.

\section*{Additional information}

\textbf{Competing financial interests:} The authors declare no competing financial interests.

%\printbibliography

\newpage
\ \\
\newpage
\section{SUPPLEMENTARY MATERIALS}

This part describes THE Supplementary Materials.

\bigskip

\bigskip

\section*{Variation of vortex state annihilation field with dot size}

\begin{figure}[h]
 \includegraphics[width=\linewidth]{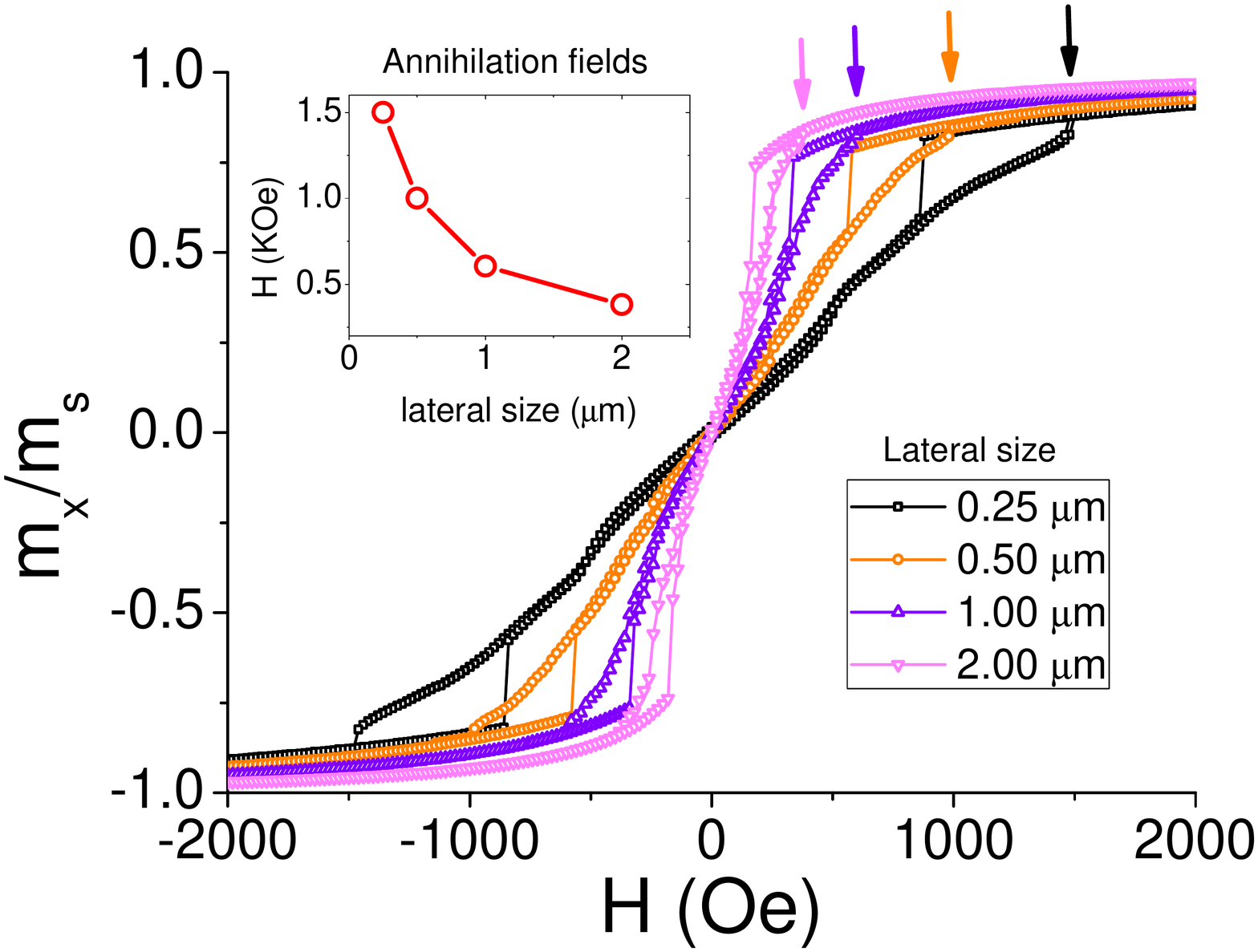}
 \caption{Simulated magnetization loops in \textbf{$30nm$ thick Py equilateral} triangular dots of different lateral size. Insert shows how vortex anihillation field changes with lateral size}
 \label{fig:hist_size}
\end{figure}

When thinking about creation of real devices one could wonder, how small could triangular dots be in practice to be able to carry edge spin waves in reasonably small bias fields.
Related point to consider when dealing with small dots is that the vortex state gets more difficult to annihilate, and threfore higher fields are necessary to reach saturation. Such change in vortex annihilation field is presented in the simulated hysteresis loops shown and analysed in Figure \ref{fig:hist_size}.

\section*{Edge domain wall profiles in triangles and rectangles}

In the main text we consider the case of patterned magnetic elements - equilateral triangular magnetic dots with sides of length 2000 nm and thickness 30 nm. In the Buckle (B) state the static bias magnetic field $H_{DC}$ is directed along the triangle base. This geometry avoids the ``butterfly'' type edge domain walls (E-DWs) which appear for fields applied completely perpendicular to the base of triangule (Y state). This type of domain wall, with a magnetic discontinuity in the middle, due to geometry also appears in rectangular or square shapes when the field is parallel to two faces and, therefore, perpendicular to the other two (figure 2 in the main text). 

%\begin{figure}[h]
%\includegraphics[width=\linewidth]{S2.pdf}
%\caption{a) Triangular and b) square dot under an applied field perpendicular to one of the faces. The lateral size is $2 \mu$m and the thickness is $30 nm$. The black streamlines and arrows represent the magnetization distribution. The color scale represents the exchange energy density.}
%\label{fig:corner_square}
%\end{figure}

Previous studies of the spin waves in rectangular magnetic elements and long strips only used in-plane high frequency fields perpendicular to the applied DC field. Then, increasing the DC field perpendicular to the edge affects the E-DW related energy so that mainly SWs are excited in the center of the strip\cite{Demidov2008}. Even if there is some splitting of the main modes into two modes that propagate closer to the sides, the perpendicular alignment of DC field and high frequency field does not excite spin waves at the edges, due to the zero torque of this excitation on the spins at the edge, which lay parallel to it, in order to reduce the magnetostatic energy outside the dot. In figure 3 in the main text we show a simulation of the corresponding effective field profiles.
%\begin{figure}[h]
%  \includegraphics[width=\linewidth]{S1.pdf}
%  \caption{Simulated profiles of the $H_x$ component of the effective magnetic field in the direction parallel to the applied external field perpendicular to the edge of rectangular element. When homogeneous RF pumping is perpenducular to the DC field, E-SWs could be excited only below some critical field when minimum of the effective field is formed close to the edge. For larger DC field excitation of the E-SWs is suppressed because torque is applied mainly inside the strip (at the bottom, the black arrow represents the DC effective field direction, and the red arrow the high frequency field direction).}
%  \label{fig:Eff_fieldprofiles}
%\end{figure}

\section*{Spectrum of edge spin waves under DC field at different angles}

\begin{figure}[h]
  \includegraphics[width=\linewidth]{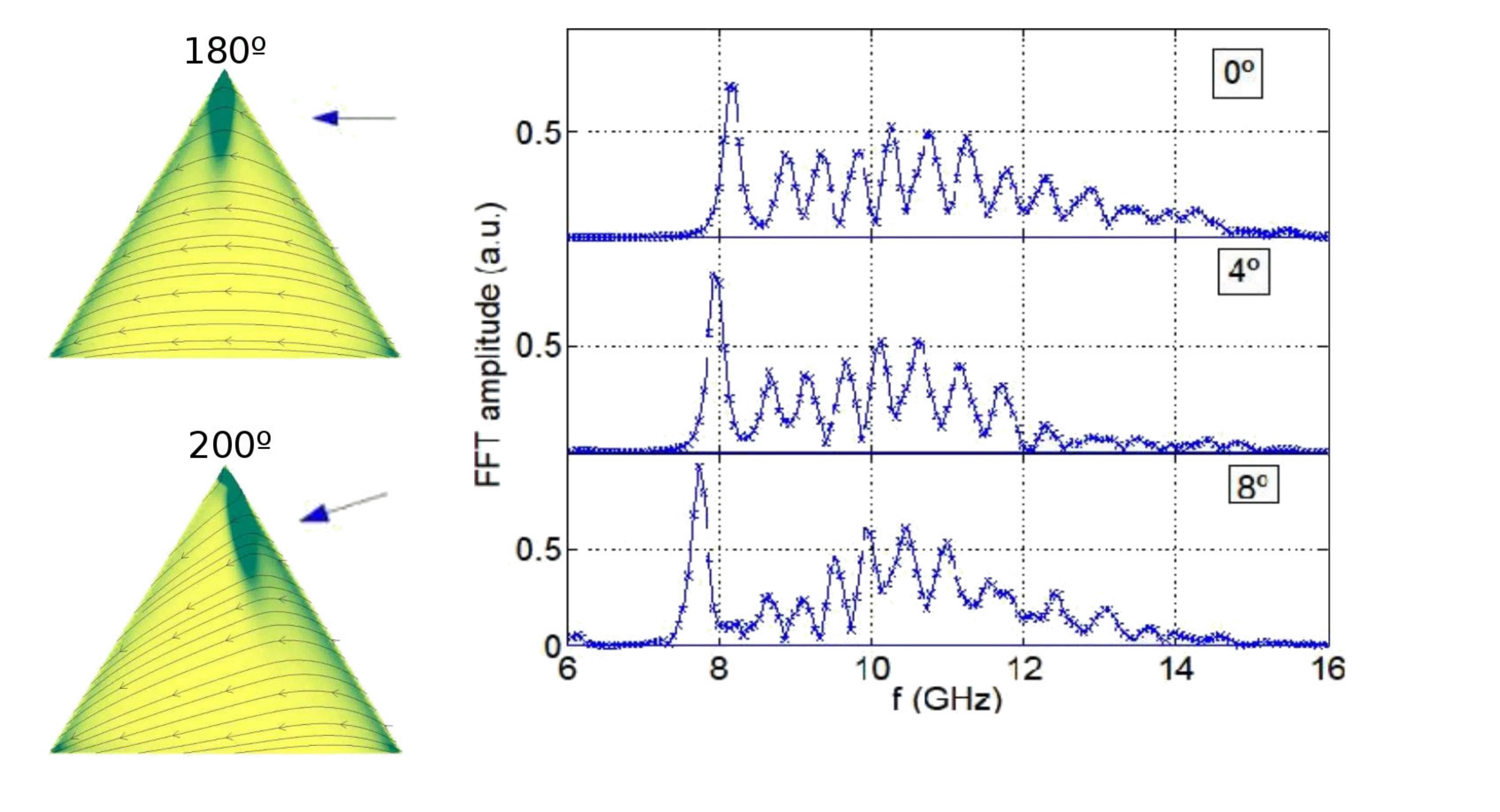}
  \caption{Left panels show the exchange energy distribution for a horizontal (top) applied field and for a 20º inclined with respect to the horizontal field (bottom). The right panel shows how eigenfrequencies are shifted even for slightly different applied field angles.}
  \label{fig:inclined}
\end{figure}

Figure \ref{fig:inclined} shows the amplitude of the FFT spectra of a 1$\mu$m triangular dot after being excited with a short gaussian field pulse. The different spectra are simulations with the DC field applied at different angles with respect to the horizontal, 0, 4 and 8 degrees. The spectra shift in frequencies (not much, but still a noticeable change) with the angle. This angle controls the magnetic state, passing from a B state at 0$^\circ$ to a Y state at 90$^\circ$, with intermediate states in between (left part, showing the exchange energy density for 0 and 20$^\circ$). If a fine tuning of the eigenmode frequencies is needed, a field perpendicular to the main DC field, in order to produce a tilted field, can do the job.

\section*{Influence of magnetic field on formation of edge domain walls in smaller dots}

\begin{figure}[h]
\centering
\includegraphics[width=\linewidth]{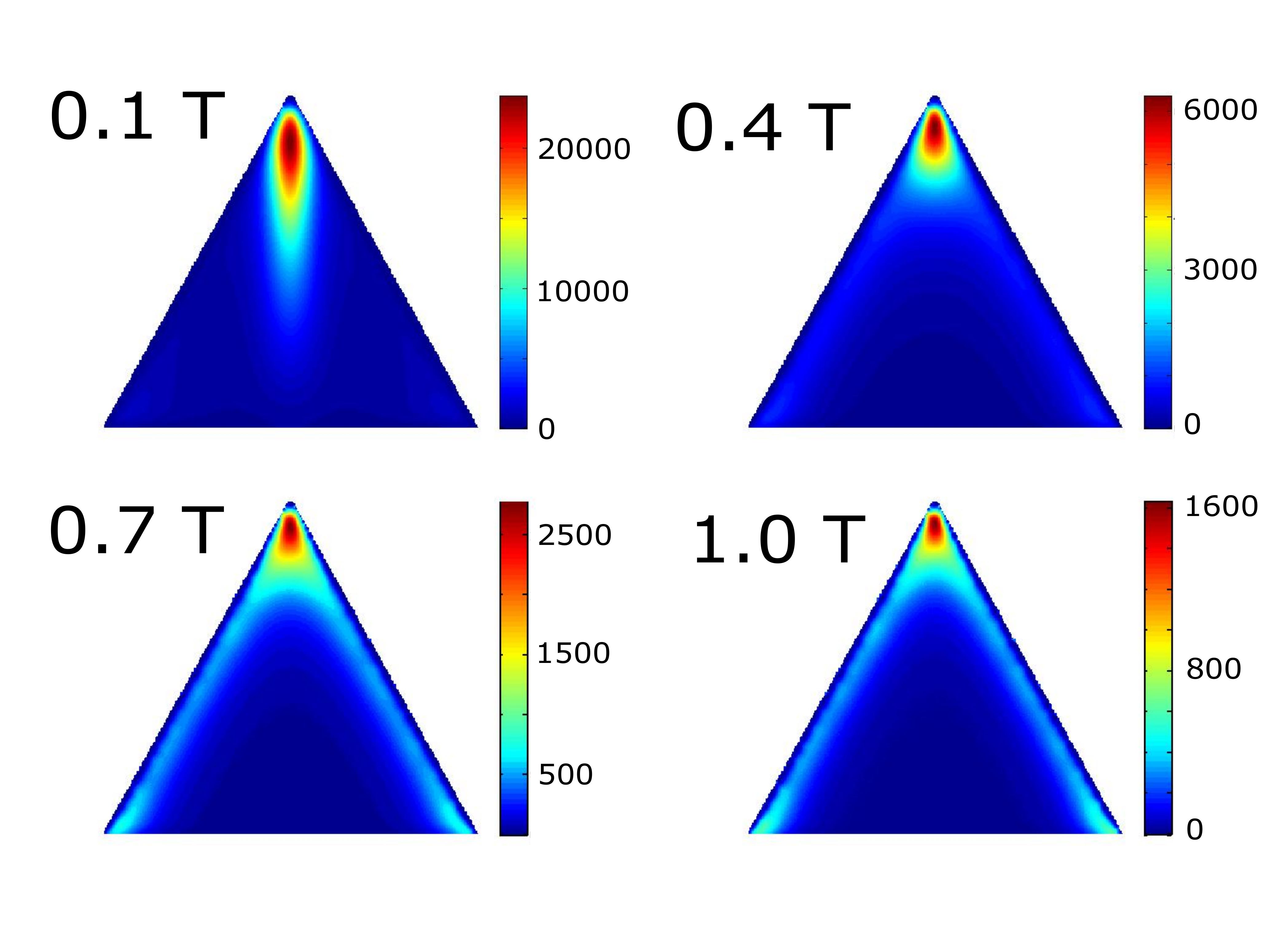}
\caption{Exchange energy density in a 0.25 $\mu$m triangular dot for different applied fields, indicated in T next to each dot. The color scale represents the exchange energy density, in J/m$^3$.}
\label{fig:exch_vs_size}
\end{figure}

The size of the triangles is important for edge spin wave transmission. In the limit of large dots, the main hurdle to overcome is wave dissipation due to damping, whereas for smaller dots it is due to the overlap that naturally occurs between the exchange channels when the size of the dot is decreased. A similar situation occurs always, for example, in the top vertex of a dot in the B state, with the field applied parallel to the base, opposite to this vertex. There, at the vertex, both exchange channels meet and become a unique magnetic charge. The only option to keep the exchange channels separate in a smaller dot is to apply a stronger magnetic field, to confine the channels closer to the edges. As this is done, the excess of exchange energy stored in the channel decreases, as shown in  Figure \ref{fig:exch_vs_size}.

\section*{Triangular dots under an exchange bias field}

\begin{figure}[h]
\centering
\includegraphics[width=\linewidth]{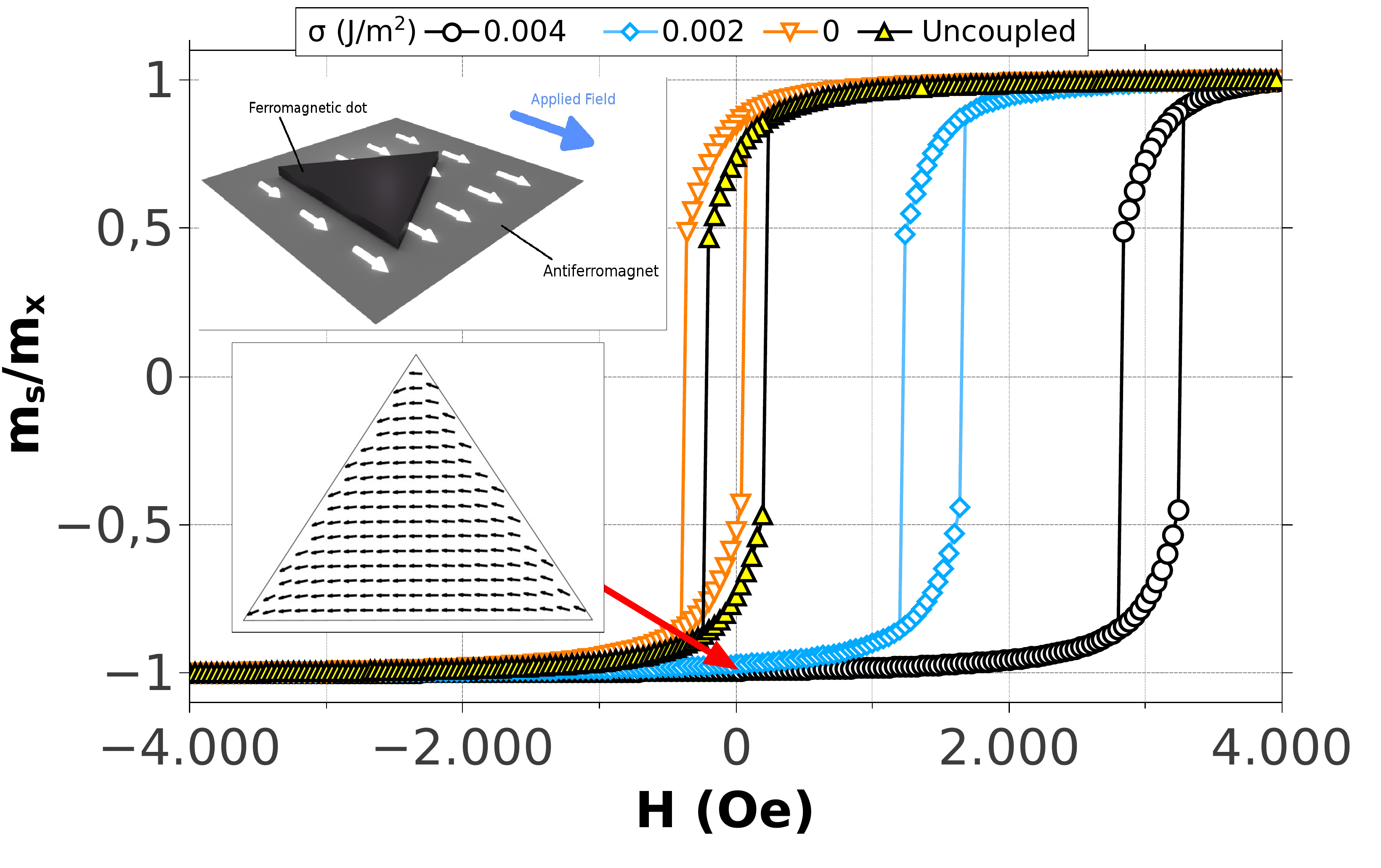}
\caption{Simulated hysteresis loops for a triangular dot in the B state with different values of exchange bias due to the coupling to an antiferromagnet.}
\label{fig:antiferro}
\end{figure}

As mentioned in the main text, it is not necessary to have an external field which saturates the sample to benefit from the properties of the B or Y states. Figure \ref{fig:antiferro} shows simulations of hysteresis loops of a 2 $\mu$m long, 30 nm thick triangular dot placed on top of a layer of fixed spins, which represents an antiferromagnet. Different exchange couplings between the two layers shift the hysteresis cycle to higher or lower fields, so that be B or the Y state could be achieved at zero external field using this method. The saturated state, with edge domain walls ready for propagating spin waves can be achieved in the absence of an external field.

\end{document}